\documentstyle[12pt]{article}
\makeatletter
\newdimen\hoogte    \hoogte=12pt    
\newdimen\breedte   \breedte=14pt   
\newdimen\dikte     \dikte=0.5pt    
\def\beginYoung{
       \begingroup
       \def\vr{\vrule height0.8\hoogte width\dikte depth 0.2\hoogte}
       \def\fbox##1{\vbox{\offinterlineskip
                    \hrule height\dikte
                    \hbox to \breedte{\vr\hfill##1\hfill\vr}
                    \hrule height\dikte}}
       \vbox\bgroup \offinterlineskip \tabskip=-\dikte \lineskip=-\dikte
            \halign\bgroup &\fbox{##\unskip}\unskip  \crcr }
\def\End@Young{\egroup\egroup\endgroup}

\makeatother

\newcommand{\beq}{\begin{equation}}
\newcommand{\eeq}{\end{equation}}
\newcommand{\bdm}{\begin{displaymath}}
\newcommand{\edm}{\end{displaymath}}
\newcommand{\beqr}{\begin{eqnarray}}
\newcommand{\eeqr}{\end{eqnarray}}
\begin{document}
\title{The generating function for a particular class of characters of $SU(n)$}
\author{W. Garc\'{\i}a Fuertes, A. M. Perelomov\footnote{On leave of absence from the Institute for Theoretical and Experimental Physics, 117259, Moscow, Russia. Current E-mail address: perelomo@dftuz.unizar.es}\\ {\small\em Departamento de F\'{\i}sica, Facultad de Ciencias, Universidad de
Oviedo, E-33007 Oviedo, Spain}} 
\date{}
\maketitle
\begin{abstract}
We compute the generating function for the characters of the irreducible representations of $SU(n)$ whose associated Young diagrams have only two rows with the same number of boxes. The result is given by formulae (\ref{ab1}), (\ref{eq:3}), (\ref{ab3})-(\ref{ab4}) and is a rational determinantal expression in which both the numerator and the denominator have a simple structure when expressed in terms of Schur polynomials.
\end{abstract}
\section{Introduction}
Among the integrable quantum mechanical systems known to date, those related to the root systems of finite dimensional simple Lie algebras form a prominent class \cite{op83}. They constitute, in particular, a natural framework to extend many classical systems of ortogonal polynomials to the case of several independent variables. A much studied example is the trigonometric Calogero-Sutherland model related to $A_{n-1}$ \cite{ca71}, \cite{su72}, whose eigenfunctions provide a natural generalization to $n$ variables of the Gegenbauer polynomials. A number of properties of these generalized Gegenbauer polynomials are known \cite{pe98a}, \cite{prz98},\cite{pe99}, \cite{pe00}. The polynomials depend on a continuous parameter $\kappa$, which is related to the coupling constant in the Hamiltonian, and are determined by $n-1$ quantum numbers. Several special values for these quantities are interesting, among which we mention two: first, when the $\kappa$ parameter goes to unity, the generalized Gegenbauer polynomials converge to the characters of the irreducible representations of $SU(n)$; second, when only the first quantum number is non-vanishing, the polynomials become those of Jack \cite{ja70}.

In the task of deepening our understanding of the properties of the generalized Gegenbauer polynomials, the computation of the generating function would be one important milestone. This is known only for the simplest $A_1$ and $A_2$ cases \cite{pe00}. Nevertheless, in some circumstances one can extract useful information from the generating function of some particular subsets of the whole system of ortogonal polynomials. The generating function of Jack polynomials, for instance, has been recently used as the starting point of an inductive proof of the structure of the derivatives of the generalized Gegenbauer polynomials \cite{fp01}. This function is a simple deformation of the generating function of the characters of the irreducible representations of $SU(n)$ obtained by taking $\kappa=1$ in the Jack polynomials. This shows how the knowledge of the generating function of some specific classes of irreducible characters of $SU(n)$ can be a valuable clue for studying the corresponding generating function for the generalized Gegenbauer polynomials. In this spirit, the purpose of this article is to compute the generating function of a subset of irreducible characters of $SU(n)$ which represents the immediate next step in complexity starting from the characters related to Jack polynomials.
\section{Computation of the generating function}
As stated in the introduction, our goal is to compute
\beq
F(t,z_i)=\sum_{k=0}^\infty P_k(z_j)\,t^k \label{eq:1}
\eeq
where $P_k(z_j)$ is the character of the irreducible representation of $SU(n)$ whose associated Young diagram has only two rows of length $k$, i.e.
\beq
P_k(z_j)=\chi_{k,k,0,\ldots,0}(z_j).
\eeq
We explain the notation. We use $\chi_{k_1,k_2,\ldots,k_n}$ to designate the character of the irreducible representation of $SU(n)$ with Young diagram containing $k_i$ boxes in the $i$-th row; $z_j$ is the $j$-th elementary symmetric polynomial in the coordinates $x_1,x_2,\ldots,x_n$ of the maximal torus of $SU(n)$:
\beq
z_j=\sum_{i_1<i_2<\ldots <i_j}x_{i_1}x_{i_2}\ldots x_{i_j}.
\eeq
There exists a simple relation between the $\chi$-symbols and the generalized Gegenbauer polynomials for $\kappa=1$, namely
\beq
\chi_{k_1,k_2,\ldots,k_n}=z_n^{k_n}\,P^1_{k_1-k_2,k_2-k_3,\ldots,k_{n-1}-k_n}. 
\eeq
Finally, we give two convenient formulae for computing the characters. First, directly in terms of the $x_j$, we have the Weyl character formula \cite{we46}
\beq
\chi_{k_1,k_2,\ldots,k_n}(x_j)=\frac{1}{\Delta}|x^{n+k_1-1},x^{n+k_2-2},\ldots,x^{k_n}|,\label{eq:2}
\eeq
in which the shorthand notation
\beq
|x^{\lambda_1},x^{\lambda_2},\cdots,x^{\lambda_n}|=\left|\begin{array}{cccc}
x_1^{\lambda_1}&x_1^{\lambda_2}&\cdots&x_1^{\lambda_n}\\
x_2^{\lambda_1}&x_2^{\lambda_2}&\cdots&x_2^{\lambda_n}\\
\vdots&\vdots&\ddots&\vdots\\
x_n^{\lambda_1}&x_n^{\lambda_2}&\cdots&x_n^{\lambda_n}\end{array}\right|
\eeq
is used, and the denominator is the Vandermonde determinant
\beq
\Delta=|x^{n-1},x^{n-2},\ldots,1|=\prod_{i<j}\,(x_i-x_j). \label{eq:vand}
\eeq
And second, as functions of $z_j$, they can be expressed through the second Giambelli identity \cite{ma95}: if $(l_1,l_2,\ldots,l_m)$ is the conjugate partition to $(k_1,k_2,\ldots,k_n)$,
\beq
\chi_{k_1,k_2,\ldots,k_n}(z_j)=\left|\begin{array}{ccccc}
z_{l_1}&z_{l_1+1}&z_{l_1+2}&\cdots&z_{l_1+m-1}\\
z_{l_2-1}&z_{l_2}&z_{l_2+1}&\cdots&z_{l_2+m-2}\\
\vdots&\vdots&\vdots&\ddots&\vdots\\
z_{l_m-m+1}&\cdots&\cdots&\cdots&z_{l_m}
\end{array}\right|
\eeq
where it is understood that $z_0=1$ and $z_j=0$ if $j>n$ or $j<0$.

After these preliminaries, we turn back to (\ref{eq:1}). According to (\ref{eq:2}), we can write
\beq
F=\frac{1}{\Delta}\sum_{k=0}^\infty(\sum_{\sigma\in S_n}{\rm sgn}(\sigma)x_{\sigma(1)}^{n-1+k}x_{\sigma(2)}^{n-2+k}x_{\sigma(3)}^{n-3}\cdots x_{\sigma(n-1)}^{1}x_{\sigma(n)}^0)t^k .
\eeq
If we interchange the summatories, we get an alternating sum of simple geometric progressions, thus
\beq
F=\frac{1}{\Delta}\sum_{\sigma\in S_n}\frac{x_{\sigma(1)}^{n-1}x_{\sigma(2)}^{n-2}\cdots x_{\sigma(n-1)}^{1}x_{\sigma(n)}^0}{1-t x_{\sigma(1)} x_{\sigma(2)}},
\eeq
which we will write as
\beq
F=\frac{g}{f}\label{ab1}
\eeq
with
\begin{eqnarray}
f&=&\prod_{i<j}(1-t x_i x_j)\label{eq:7}\\
g&=&\frac{1}{\Delta}\sum_{\sigma\in S_n}{\rm sgn}(\sigma)x_{\sigma(1)}^{n-1}x_{\sigma(2)}^{n-2}\cdots x_{\sigma(n-1)}\frac{f}{1-t x_{\sigma(1)} x_{\sigma(2)}}\label{eq:7b}.
\end{eqnarray}
$f$ and $g$ are polynomials in $t$ of respective degrees $N=\frac{1}{2}n(n-1)$ and $N-1$. It is obvious that the coefficients of $f$ are homogeneous symmetric polynomials in $x_i$ over the integers. As $F$ is a series of polynomials of the same kind, the statement turns out to be also true for $g$. Our next task is to compute $f$ and $g$ in closed form; we would like, in particular, to express their coefficients in the most natural basis for homogeneous symmetric polynomials in the present context: the irreducible characters of $SU(n)$, or Schur polynomials.

In fact, the result for $f$ is known: it was obtained by Weyl in the course of his computation of the characters of the symplectic groups. The Weyl result is \cite{we46}
\beq
f=\frac{1}{\Delta}|x^{n-1},x^{n-2}+x^n,x^{n-3}+x^{n+1},\ldots,1+x^{2n-2}| {\Large |}_{x->\sqrt{t} x}.\label{eq:3}
\eeq
Given this expression, we can obtain the coefficients in $t$ in
\beq
f=1-Z_1\,t+Z_2\,t^2-Z_3\,t^3+\cdots +(-1)^N Z_N\,t^N \label{eq:3p}
\eeq
by expanding the determinant in the numerator of (\ref{eq:3}) in such a way that all terms have monomial columns and collecting terms of the same order in $t$. This gives $Z_d$ as a sum of determinantal quotients of the form
\beq
z_\lambda=\frac{1}{\Delta}|x^{n+\lambda_r},x^{n+\lambda_{r-1}},\ldots,x^{n+\lambda_1},x^{n-1},\ldots,\widehat{x^{n-(\lambda_1-2)}},\ldots,\widehat{x^{n-(\lambda_r-2)}},\ldots,1| {\Large |}_{x->\sqrt{t} x}\label{eq:w1}
\eeq
where the hat over a term means that the term is absent and the sum is is extended to all possible combinations such that 
\begin{eqnarray}
n-2\geq \lambda_r >\lambda_{r-1}>,\ldots,>\lambda_1&\geq& 0\nonumber\\ \lambda_1+\ldots+\lambda_r&=&d-r.
\end{eqnarray} 
From (\ref{eq:w1}), the Weyl character formula gives $z_\lambda=\chi_{k_1,k_2,\ldots,k_n}$ with
\beq
k_j=\left\{ \begin{array}{ll}
j+\lambda_{r-j+1} & 1\leq j\leq r \\
r & r+1\leq j\leq r+\lambda_1+1 \\
r-k & r+\lambda_k+3-k\leq j \leq r+\lambda_{k+1}+1-k,\ \ k=1,2,\ldots,r-1 \\
0 & j\geq \lambda_r+3
\end{array} \right.
\eeq
A closer inspection shows that this structure corresponds to a Young diagram of rank $r$ with $j+\lambda_{r-j+1}$ boxes in the $j$-th row and in which the number of rows with number of boxes greater or equal to $j$ is $1+j+\lambda_{r-j+1}$, $j=1,2,\ldots,r$. Therefore, the Young diagram associated to $\chi_{k_1,k_2,\ldots,k_n}$ is $(\lambda_r,\lambda_{r-1},\ldots,\lambda_1 |\lambda_r+1,\lambda_{r-1}+1,\ldots,\lambda_1+1)$ in Frobenius notation, see \cite{ma95}, and, by identifying each diagram with its associated $SU(n)$ character, we can write
\beq
Z_d=\sum_{{\cal P}_d} (\beta_1-1,\beta_2-1,\ldots,\beta_r-1|\beta_1,\beta_2,\ldots,\beta_r)\label{eq:3pp}
\eeq
where ${\cal P}_d$ is the set of partitions of $d$ such that $\beta_1>\beta_2>\cdots>\beta_r\geq 1$.

On the other hand, we can show that $g$ is given by a determinantal expression very similar to (\ref{eq:3}):
\beq
g=\frac{1}{\Delta}|x^{n-1},x^{n-2},x^{n-3},x^{n-4}+x^n,x^{n-5}+x^{n+1},\ldots,1+x^{2n-4}| {\Large |}_{x->\sqrt{t} x}. \label{eq:5}
\eeq
To see this, it is convenient to rescale temporarily $t$ to the unity, to write $g=\frac{Q_1}{\Delta}$ and to consider first the case $x_1=x_2^{-1}$. From (\ref{eq:7b}) we get
\begin{eqnarray}
Q_1|_{x_1=x_2^{-1}}&=&[(x_1^{n-1} x_2^{n-2}x_3^{n-3}\cdots x_{n-1}+{\rm perm})-(x_2^{n-1} x_1^{n-2}x_3^{n-3}\cdots x_{n-1}+{\rm perm})]\cdot\nonumber\\
&\cdot &\{  [(1-x_1 x_3)\cdots(1-x_1 x_n)][(1-x_1^{-1} x_3)\cdots(1-x_1^{-1} x_n)]\prod_{j,k=3}^n(1-x_j x_k)\} =\nonumber\\
&=&(x_1-x_2)[x_3^{n-3}x_4^{n-4}\cdots x_{n-1}+{\rm perm}]\cdot\nonumber\\
&\cdot &\{  [(1-x_1 x_3)\cdots(1-x_1 x_n)][(1-x_1^{-1} x_3)\cdots(1-x_1^{-1} x_n)]\prod_{j,k=3}^n(1-x_j x_k)\}\label{eq:q1}
\end{eqnarray}
where ``perm" refers to the permutations of the powers of $x_3,x_4,\ldots,x_{n-1}$ including signature. By taking $x_k^{n-2}$ as a common factor in the $k$ row of the numerator of (\ref{eq:5}) for $k=1$ to $n$, we write:
\begin{eqnarray}
Q_2&=&|x^{n-1},x^{n-2},x^{n-3},x^{n-4}+x^n,x^{n-5}+x^{n+1},\ldots,1+x^{2n-4}|\nonumber\\
&=&(x_1 x_2\cdots x_n)^{n-2}|x,1,y,y^2,\ldots,y^{n-2}|
\end{eqnarray}
with $y_j=x_j+x_j^{-1}$. If we take $x_1=x_2^{-1}$ in this expression and subtract in the determinant the second from the first row, we obtain through the Vandermonde formula (\ref{eq:vand})
\beq
Q_2|_{x_1=x_2^{-1}}=(-1)^{\frac{(n-1)(n-2)}{2}}(x_1-x_2)(x_3 x_4\cdots x_n)^{n-2}(y_2-y_3)\cdots(y_2-y_n)\prod_{\stackrel{i,j=3}{i<j}}^n (y_i-y_j)\label{eq:w2}.
\eeq
The use of the identities
\begin{eqnarray}
(1-x_2^{-1}x_j) (1-x_2 x_j)&=&-x_j(y_2-y_j)\nonumber\\
y_k-y_j&=&-(x_k x_j)^{-1}(x_k-x_j)(1-x_k x_j)
\end{eqnarray}
in (\ref{eq:w2}) transforms the right side of this equation in exactly the last member of (\ref{eq:q1}); therefore $Q_1|_{x_1=x_2^{-1}}=Q_2|_{x_1=x_2^{-1}}$. As $\frac{Q_1}{\Delta}$ and $\frac{Q_2}{\Delta}$ are symmetric polynomials, this implies that $\frac{Q_1}{\Delta}-\frac{Q_2}{\Delta}= f P$ where $P$ is a symmetric polynomial, but from (\ref{eq:7}) and the definitions of $Q_1$ and $Q_2$, one can easily check that the total degree of the left side of this equation is necessarily lower than the total degree of $f$, so that $P=0$. This concludes the proof of (\ref{eq:5}).

If we now write
\beq
g=\sum_{k=0}^{N-1}G_k t^k \label{ab3}
\eeq
a computation completely analogous to that leading from (\ref{eq:3}) to (\ref{eq:3pp}) allows us to write $G_k$ as follows 
\beq
G_k=\sum_{{\cal Q}_{k+j,j}}(-1)^{k+j} (\beta_1-3,\beta_2-3,\ldots,\beta_j-3|\beta_1,\beta_2,\ldots,\beta_j),
\eeq
where ${\cal Q}_{k+j,j}$ is the set of partitions of $k+j$ with $j$ terms and satisfying $\beta_1>\beta_2>\cdots\beta_j\geq 3$.
From (\ref{eq:3}) and (\ref{eq:5}), we give the final formula for the desired generating function:
\beq
F=\frac{|x^{n-1},x^{n-2},x^{n-3},x^{n-4}+x^n,x^{n-5}+x^{n+1},\ldots,1+x^{2n-4}|}{|x^{n-1},x^{n-2}+x^n,x^{n-3}+x^{n+1},\ldots,1+x^{2n-2}|}{\Large|}_{x->\sqrt{t} x}.\label{ab4}
\eeq
\section{Differential equations for $f$ and $g$}
We will deduce in this section two differential equations satisfied by $f$ and $g$. These equations can be taken as the basis for an alternative approach for the computation of these quantities.

In establishing the equations, we will take advantage of two differential operators of a class introduced in \cite{ha82}, \cite{ha83}, \cite{fo49}:
\begin{eqnarray*}
D_1&=&\sum_{p=1}^n z_{p-1}\frac{\partial}{\partial z_p}\\
D_2&=&\frac{1}{2}[\sum_{p=2}^n z_{p-2}\frac{\partial}{\partial z_p}+D_1^2]
\end{eqnarray*}
The action of these operators on Schur polynomials can be most simply described in graphical terms: $D_1$ applied to the Schur polynomial $S_\lambda$ gives the sum of all Schur polynomials whose associated Young diagrams are that of $S_\lambda$ with one box removed. $D_2$ does similarly but, in this case, the sum is over the Young diagrams obtained by removing two boxes not in the same row in all possible ways.

From (\ref{eq:7}), we get
\beq
\ln f=-\sum_{i<j}(x_i x_j+\frac{1}{2}x_i^2 x_j^2+\frac{1}{3}x_i^3 x_j^3+\cdots)
\eeq
and, therefore
\beq
f=\exp\{-\sum_{k=1}^\infty \frac{m_k}{k}t^k\},\ \ \ \ \ \ \ m_k=\sum_{i<j}^n x_i^k x_j^k
\eeq
or, alternatively
\beq
f=\exp\{-\frac{1}{2}\sum_{k=1}^\infty\frac{p_k^2-p_{2k}}{k}t^k\},\ \ \ \ \ \ \ p_k=\sum_{i=1}^n x_i^k \label{eq:8}.
\eeq
This gives ${\displaystyle \frac{\partial f}{\partial p_1}=-p_1 t f}$. But $p_1=z_1$  and ${\displaystyle\frac{\partial}{\partial p_1}=D_1}$ \cite{fo49}, \cite{ma95}, hence
\beq
D_1 f=-z_1 t f
\eeq
or, using (\ref{eq:3p})
\beq
D_1 Z_j=z_1 Z_{j-1}.
\eeq
To find the differential equation for $g$, we use (\ref{eq:8}) to write
\beq
(\frac{1}{2}\frac{\partial^2}{\partial p_1^2}-\frac{\partial}{\partial p_2})f=[\frac{1}{2}t^2(p_1^2+p_2)-t] f
\eeq
and, as $z_2=\frac{1}{2}(p_1^2-p_2)$ and ${\displaystyle D_2=\frac{1}{2}\frac{\partial^2}{\partial p_1^2}-\frac{\partial}{\partial p_2}}$ \cite{fo49}, \cite{ma95}, we conclude that
\beq
D_2 f=[t^2(z_1^2-z_2)-t]f.
\eeq
Now, 
\beq
D_2 g=D_2(f F)=(D_2f)F+f(D_2F)+(D_1f)(D_1F).
\eeq
The first term is proportional to $g$. The second too: as the diagrams in $F$ consist only of two identical lines, we get
\beq
D_2 F=t F.
\eeq
The third term gives a contribution proportional to $D_1 g$, because
\beq
D_1 g= D_1(f F)=-t z_1 g+f D_1 F.
\eeq
The differential equation is therefore
\beq
D_2g+t z_1 D_1 g+t^2 z_2 g=0
\eeq
or, alternatively
\beq
D_2 G_j+z_1 D_1 G_{j-1}+z_2 G_{j-2}=0.
\eeq
  \\\\\\
{\Large\bf Acnowledgments}\\\\
A.M.P. would like to express his gratitude to the Department of Physics of the University of Oviedo for the hospitality during  his stay as a Visiting Professor. The work of W.G.F. has been partially supported by grant BFM 2000 0357 (DGICYT, Spain).\\

\end{document}